\documentclass{ws-procs975x65}

\begin{document}

\title{Improved Torsion Pendulum for Ground Testing of LISA Displacement Sensors}

\author{L.~Carbone, R.~Dolesi, C.D.~Hoyle, M.~Hueller, S.~Vitale, and W.J.~Weber}

\address{Dipartimento di Fisica, Universit\'a di Trento, 
38050 Povo (TN), Italy\\E-mail: hoyle@science.unitn.it}

\author{A. Cavalleri}

\address{Centro Fisica degli Stati Aggregati, 
38050 Povo (TN), Italy}  

\maketitle

\abstracts{
We discuss a new torsion pendulum design for ground testing of prototype LISA (Laser Interferometer Space Antenna) displacement sensors. This new design is directly sensitive to net forces and therefore provides a more representative test of the noisy forces and parasitic stiffnesses acting on the test mass as compared to previous ground-based experiments. We also discuss a specific application to the measurement of thermal gradient effects.}

\section{Background and Motivation}\label{subsec:back}
%\subsection{General}
LISA\cite{pb1} aims to measure gravitational radiation in the low frequency band 0.1-100~mHz with a constellation of three satellites containing six ``free-falling'' test masses, each of which is one end mirror of an interferometer beam that forms the side of an equilateral triangle of length 5$\times$10$^6$~km. The interferometric measurement of the relative test mass displacement gives the gravitational wave strain. The strain sensitivity is predicted to be limited at low frequencies by stray forces that produce spurious acceleration of the test masses. In order to achieve the sensitivity goals, each test mass must not be perturbed from geodetic motion to within $S^{1/2}_{a_n}\leq 3\times10^{-15}$~m/s$^2$/Hz$^{1/2}$ along its sensitive interferometer axis, $x$. For the envisioned 2~kg test masses, this represents a stray force noise of $S^{1/2}_{f_n}\leq 6\times10^{-15}$~N/Hz$^{1/2}$.

A satellite shield, kept centered around the test masses by a six degree of freedom displacement sensor, will protect the test masses from many environmental disturbances. The noisy force acting on a test mass can be written as
\begin{equation}
f_n\,=\,f_{str}\,+\,k_p\left(x_n\,+\,\frac{F_{str}}{M\omega^2_{DF}}\right),
\end{equation}
where $f_{str}$ are spurious forces acting on the test mass, $k_p$ is any spring-like coupling (or ``stiffness'') between the satellite and test mass, $x_n$ is the noise in the displacement sensor itself, $F_{str}$ represents the forces acting on the satellite, $M$ is the entire satellite mass, and $\omega^2_{DF}$ represents the finite gain of the ``drag free'' control loop that keeps the satellite centered about the test mass. Thus we see that it is important that the displacement sensor itself is not a source of spurious forces or residual coupling between the satellite and test mass.

A prototype displacement capacitive sensor that meets the LISA goals in terms of displacement sensitivity, noise, and parasitic stiffness has been designed.\cite{sens} However, due to the extreme level of disturbance isolation needed for LISA, it is important to unexpected sources of force noise or stiffness arising in the sensor. The space-borne LISA Pathfinder mission\cite{LTP}$^,$\cite{ST7} will attempt to prove the performance to within a factor of ten of the LISA goal, in a restricted band. However, it is additionally desirable to have a ground testing program to build confidence in the design before the flight test.

Torsion pendulums have proven to be useful instruments for measuring the stray forces and stiffnesses arising in LISA-like displacement sensors, as well as characterization of specific sources of stray forces.\cite{oldpend}$^,$\cite{prl}$^,$\cite{cqgpend} However, all previous results have been sensitive only to net torques on the test mass. The characterization of the random forces thus depends on the conversion through an effective arm-length. As the effective arm-length varies depending on the type of disturbance, it is desirable to have a configuration sensitive directly to net forces along the sensitive $x$ axis.  

\section{The 4-Mass Design}\label{design}
\begin{figure}[t]
\epsfxsize=2.2in % will enlarge or reduce the postscript figures based on the xsize
\hfil\epsfbox{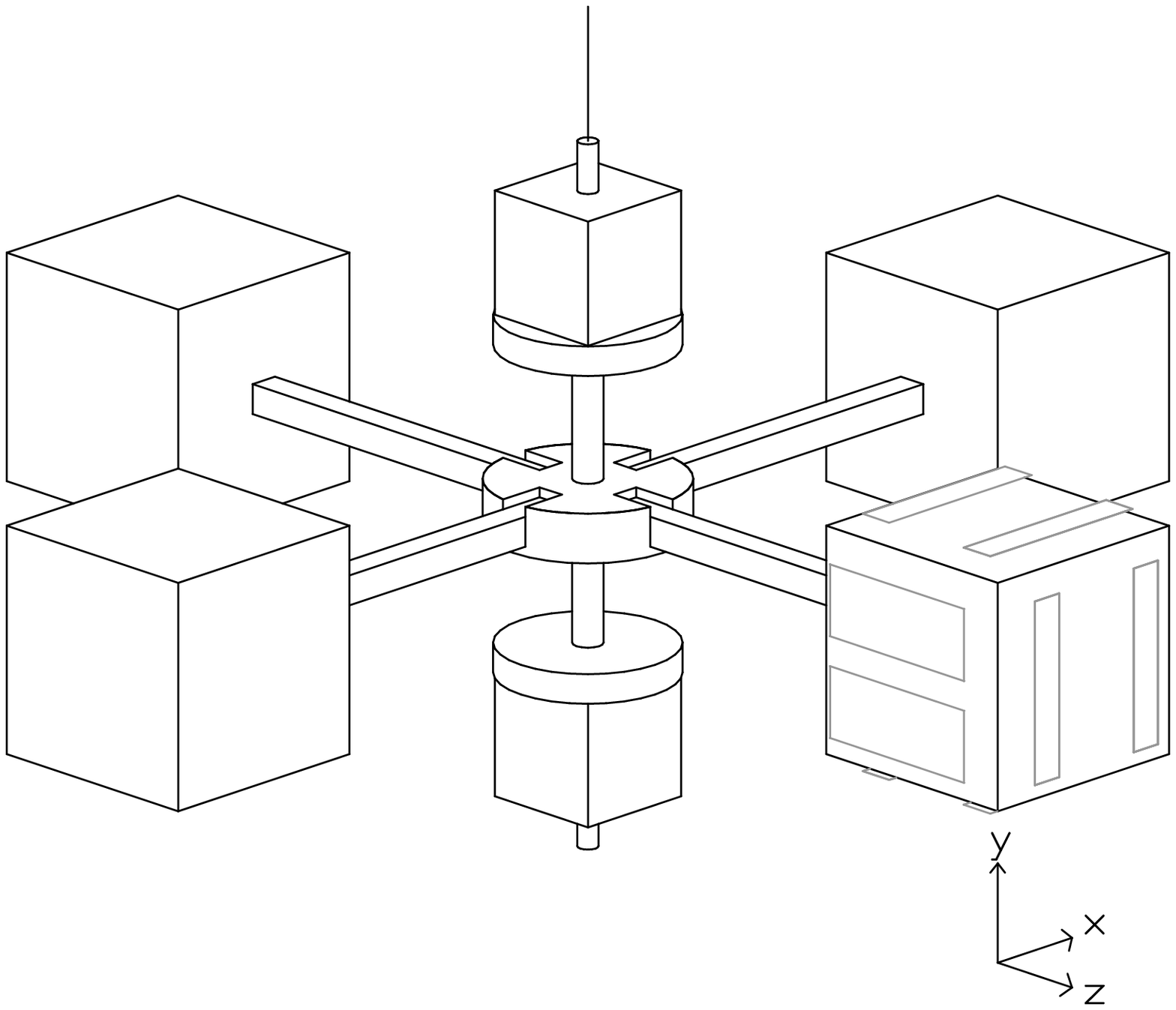}\hfil
\epsfxsize=2.6in
\hfil\epsfbox{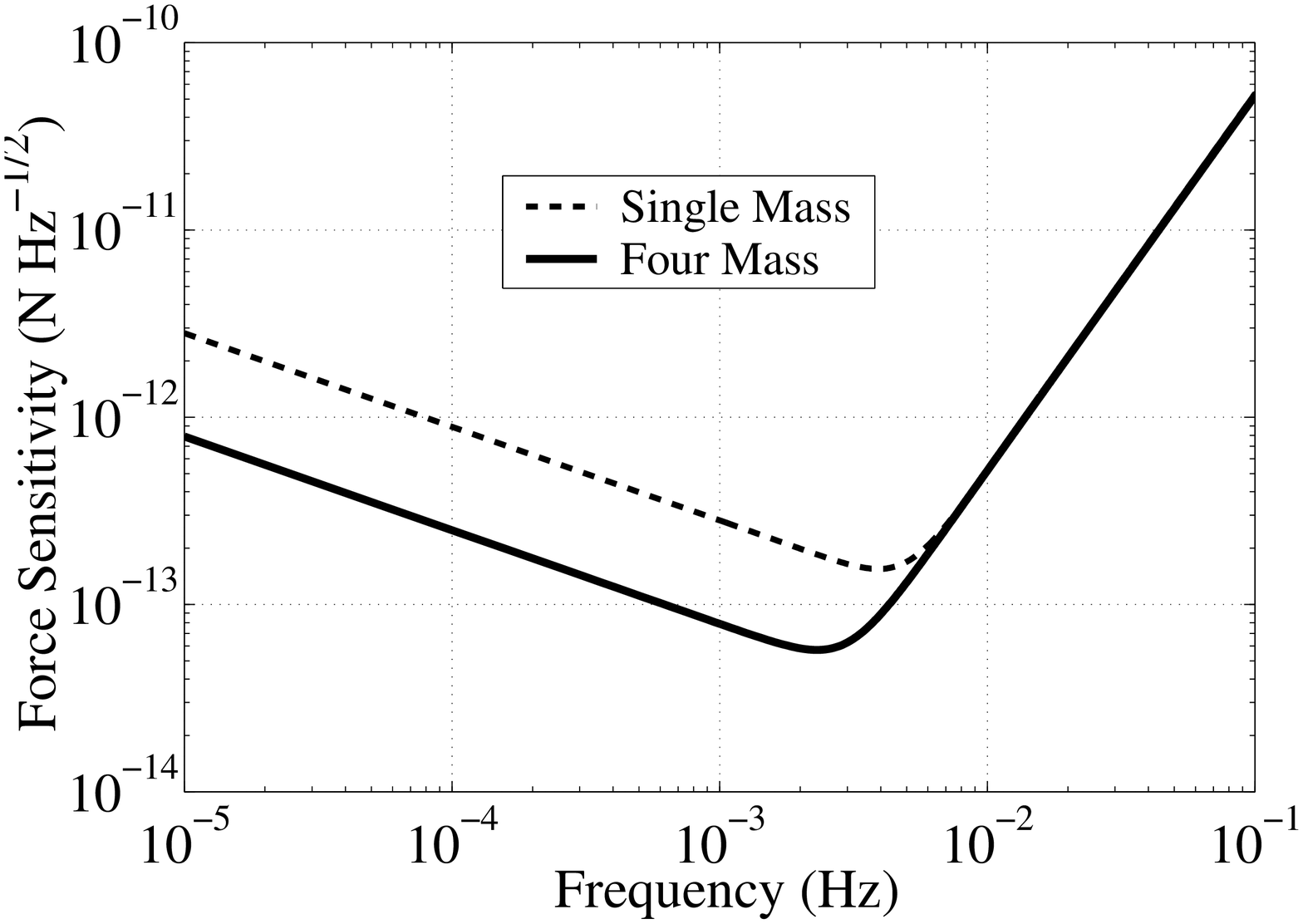}\hfil % postscript image file name
\caption{Left: Conception of the four-fold symmetric torsion pendulum. The centers of the four hollow, cubic test masses are displaced 11~cm from the torsion fiber axis. The smaller cubes displaced along the $y$-axis (vertical) serve to reduce the gravitational quadrapole moment and provide a reflecting surface for an optical readout of the pendulum twist. 
The positions of the displacement sensor electrodes, outlined in gray, are shown surrounding the mass on the lower right. 
%The 12 electrodes used to measure displacement in all 6 degree of freedom are outlined in gray.
%, while those used to apply a 100~kHz, AC bias to the test mass are black. 
Right: Predicted force sensitivity of the pendulum shown on the left (using a 53~$\mu$m diameter tungsten fiber 
of length 1~m and quality factor $Q\,=\,4000$) as compared to the single mass configuration. The four-mass design is directly sensitive to net forces, while the single mass curve is interpreted from its torque sensitivity using an effective arm-length of 1~cm. \label{fig1}}
\end{figure}

In order to produce a configuration sensitive to net forces with a torsion pendulum, it is necessary to displace the test mass from the torsion fiber axis. This can be done by adopting a simple Cavendish type geometry; however, as this displacement, $R$, increases, the pendulum quickly becomes more susceptible to environmental gravity gradient noise. It is thus necessary to have a high degree of symmetry to suppress couplings to low-order multipole moments.\cite{su} Initial simulations show that a four mass design with quadrapole compensation, such as that shown in Figure~\ref{fig1}, should be substantially insensitive to gravitational disturbances. 

The instrument limit for such a pendulum depends on the thermal noise in the pendulum motion
%, $S_{N}$, 
and the displacement sensor position noise, $S_{x_n}$:
\begin{equation}
%S_{f_n}^{1/2}(\omega)=\frac{\sqrt{S_N(\omega)+\Gamma^2S_{x_n}/(R^2|F(\omega)|^2)}}{R}\approx\frac{\sqrt{4k_BT\,\Gamma/\omega Q+\Gamma^2S_{x_n}/(R^2\,|F(\omega)|^2)}}{R}, 
S_{f_n}^{1/2}(\omega)\approx\frac{\sqrt{4k_BT\,\Gamma/\omega Q+\Gamma^2S_{x_n}/(R^2\,|F(\omega)|^2)}}{R}, 
\end{equation}
where $k_B$ is Boltzmann's constant, $T$ is the temperature, $\Gamma$ is the torsional spring constant of the suspension fiber, $Q$ is the pendulum resonance quality factor, and $|F(\omega)|^2\,=\,((1-\omega^2/\omega^2_\circ)^2+1/Q^2)^{-1}$ is the pendulum's angular transfer function with resonant angular frequency $\omega_\circ$. To maximize the sensitivity we see that it is important to have a soft, high $Q$ torsion fiber as well as a large arm length. As previously stated, however, the arm-length cannot be arbitrarily large because gravitational noise sources will begin to dominate at some point. For a given fiber length, the torsional spring constant varies as $\Gamma\,\propto\,M^2$, where $M$ is the entire pendulum mass that the fiber can support, making it advantageous to make everything as light as possible. 

The right hand side of Figure~\ref{fig1} shows a realistic force sensitivity curve for the pendulum design shown on the left. This aluminum pendulum has a total mass of $\approx450$~g, necessitating a tungsten torsion fiber with $\Gamma\,=\,1\times10^{-7}$~N-m/rad. The position sensor is assumed to have the demonstrated intrinsic noise level of 0.4~nm/Hz$^{1/2}$. 

\section{Applications}\label{apps}
In addition to measuring the random force noise on the test mass along the sensitive degree of freedom, the torsion pendulum may be used to investigate, quantitatively, specific sources of disturbances that are predicted to be present in LISA. By modulating specific source terms we can determine the coupling of the test mass to the effect in question. The multiple mass design permits measurements of the net forces associated with temperature gradients, magnetic fields and field gradients, and electrostatic and gravitational interactions.   

Of particular interest is the effect of thermal gradient fluctuations on the test mass. Thermal gradients can produce a force on the test mass through several mechanisms including radiation pressure, radiometric effects, and temperature dependent outgassing. While the prototype sensors have been designed to have a high thermal conductance and thus small susceptibility to thermal gradients, and the expected temperature fluctuations aboard LISA are a small $S_{{\Delta}T}=0.1$~mK/$\sqrt{\rm Hz}$ (where ${\Delta}T$ is taken across the sensor), it is nevertheless worthwhile to demonstrate that there is not an excess of force noise induced by poorly understood outgassing or other unmodeled effects.

Initial simulations suggest that an induced temperature difference may be created across the displacement sensor with small heaters ($\sim$1~W) attached to the sensor housing. With such a an induced gradient, the radiometric effect should be dominant, applying a net force of\cite{schu} $F_{r}\approx5\times10^{-11}$~N to the test mass. The contribution from radiation pressure should be smaller,\cite{schu} $F_{rp}\approx3\times10^{-12}$~N, but the two may be distinguished by their respective temperature and pressure, $P$, dependences: $F_{r}{\,\propto\,}P/T$ and $F_{rp}{\,\propto\,}T^3$. In any case, this level of force will be definitely discernable with the torsion pendulum design outlined above by modulating the temperature gradient and observing the coherent deflection of the pendulum.

A preliminary estimate of the outgassing effect\cite{sens} predicts the level to be roughly 10\% that of the radiometric effect and nominally independent of the pressure inside the sensor. Any large, pressure independent signal coherent with the temperature gradient modulation would thus be a sign of outgassing or other spurious effects. 

\section{Conclusions and Outlook}\label{conc}
We have described a new torsion pendulum design to be used to test the level of force isolation permitted by a LISA prototype capacitive displacement sensor on ground. The new pendulum will improve upon previous ground tests because of its sensitivity to net forces along the axis sensitive to the gravitational wave strain. We are currently in the design and construction phase of this apparatus and hope to have results within the next year.

\section*{Acknowledgments}
This work was supported by ESA, INFN, and ASI.

\end{document}